\documentclass[pre,twocolumn,showpacs]{revtex4}
\usepackage{epsfig}
\usepackage{amsmath}
\usepackage{amssymb}

\begin{document}

\title{Some exact solutions of the local induction equation for motion of a vortex
in a Bose-Einstein condensate with Gaussian density profile}
\author{V. P. Ruban}
\email{ruban@itp.ac.ru}
\affiliation{Landau Institute for Theoretical Physics RAS, Moscow, Russia} 

\date{\today}

\begin{abstract}
The dynamics of a vortex filament in a trapped Bose-Einstein condensate is considered
when the equilibrium density of the condensate, in rotating with angular velocity 
${\bf\Omega}$ coordinate system, is Gaussian with a quadratic form ${\bf r}\cdot\hat D{\bf r}$. 
It is shown that equation of motion of the filament in the local induction approximation admits 
a class of exact solutions in the form of a straight moving vortex, 
${\bf R}(\beta,t)=\beta {\bf M}(t) +{\bf N}(t)$, where $\beta$ is a longitudinal parameter, and
$t$ is the time. The vortex is in touch with an ellipsoid, as it follows from the conservation laws
${\bf N}\cdot \hat D {\bf N}=C_1$ and ${\bf M}\cdot \hat D {\bf N}=C_0=0$. Equation of motion
for the tangent vector ${\bf M}(t)$ turns out to be closed, and it has the integrals 
${\bf M}\cdot \hat D {\bf M}=C_2$, $(|{\bf M}| -{\bf M}\cdot\hat G{\bf \Omega})=C$, where the matrix
$\hat G=2(\hat I \mbox{Tr\,} \hat D -\hat D)^{-1}$. Intersection of the corresponding level surfaces
determines trajectories in the phase space.

\end{abstract}


\maketitle

{\bf Introduction}.
Theoretical description  of the dynamics of a quantized vortex filament in a rotating trapped
Bose-Einstein condensate is a complicated physical problem (see, e. g., \cite{FS2001,F2009,SF2000},
and many references therein). Some simplification is possible if the rotation frequency $\Omega$
is small in comparison with a characteristic transverse frequency $\omega_\perp$ of the trap,
and the condensate itself is in the so called Thomas-Fermi regime (i. e., 
$[\mu-V_{\rm min}]\gg\hslash\omega_\perp$, where $\mu$ is the chemical potential, and $V_{\rm min}$
is the minimum value of the trap potential). Under such conditions, the unperturbed 
vortex-free  density field inside the condensate, i. e. not too closely to the Thomas-Fermi surface
$[\mu -V({\bf r})]=0$, is given with a good accuracy by expression  
$\rho_0\approx \mbox{const}\cdot[\mu -V({\bf r})]$ (it is implied that the wave function of 
the condensate follows the Gross-Pitaevskii equation). At that, a typical condensate size $\tilde R$
is much larger than a vortex core width $\xi$, and the density far away from the vortex is 
practically static (in the absence of potential excitations). Behaviour of the quantized 
vortex filament with a reasonable accuracy is described by the classical equations of the slow
hydrodynamics of an inviscid compressible fluid against a given density background.
Two technical difficulties are inherent to this problem already at its initial stage.
The first difficulty comes from the necessity to find the unperturbed vortex-free velocity field
${\bf v}_0({\bf r})$ of the condensate itself in the trap-co-rotating coordinate system,
so one has to solve a system of partial differential linear equations with non-constant coefficients,
\begin{equation}\label{stat}
{\bf v}_0=\nabla \varphi-[{\bf \Omega}\times {\bf r}], 
\qquad \nabla\cdot(\rho_0 {\bf v}_0)=0,
\end{equation}
where $\bf\Omega$ is the angular velocity vector of the trap rotation.

Let in the condensate be a single vortex with one quantum of circulation 
$\Gamma=2\pi\hslash/m_{\rm atom}$. Its dynamics in the three-dimensional (3D) space is described by 
an unknown vector function ${\bf R}(\beta, t)$, where $\beta$  is an arbitrary longitudinal 
parameter, and $t$ is the time. Equation of motion of a thin vortex filament in the classical
hydrodynamics follows from a variational principle with the Lagrangian of the form (see details
in \cite{R2001,R2016})
\begin{equation}
{\cal L}=\Gamma \oint({\bf F}({\bf R})\cdot[{\bf R}_\beta \times{\bf R}_t]) d\beta -{\cal H}\{{\bf R}\},
\end{equation}
with vector function ${\bf F}({\bf R})$ satisfying the condition
\begin{equation}
\nabla\cdot{\bf F}({\bf R})=\rho_0({\bf R}).
\end{equation}
The Hamiltonian functional ${\cal H}\{{\bf R}\}$ is the sum 
\begin{equation}
{\cal H}\{{\bf R}\}={\cal K}_\Gamma\{{\bf R}\} +\Gamma\oint ({\bf A}({\bf R})\cdot {\bf R}_\beta) d \beta,
\end{equation}
where ${\cal K}_\Gamma\{{\bf R}\}$ is the kinetic energy of the vortex itself, and
${\bf A}({\bf r})$ is a vector potential of the unperturbed (mass) density of current, i. e., 
$\rho_0{\bf v}_0=\mbox{curl}\,{\bf A}$. The following equality takes place:
\begin{equation}
{\cal K}_\Gamma=(\Gamma/2)\int_{S_\Gamma} (\rho_0{\bf v}_\Gamma\cdot d{\bf S}),
\end{equation}
where $S_\Gamma$ is a surface spanned on the vortex contour,  ${\bf v}_\Gamma$ is the self-consistent
velocity field created by the (quasi)singular vortex filament (the integration is ``cut'' at a distance 
of order $\xi$ from the vortex line). The second mentioned technical difficulty is that the given
integral is impossible to be written in a closed form in terms of the function ${\bf R}(\beta, t)$,
because the spatial non-uniformity of density does not allow an exact analytical calculation 
of ${\bf v}_\Gamma$, with a few exceptions. However, in the so called local induction approximation 
the functional ${\cal K}_\Gamma\{{\bf R}\}$ can be always expressed with a logarithmic accuracy:
\begin{equation}
{\cal K}_\Gamma\{{\bf R}\}\approx({\Gamma^2\Lambda}/{4\pi})\int \rho_0({\bf R}) |{\bf R}_\beta|d\beta,
\end{equation}
where $\Lambda=\log(\tilde R/\xi)\approx\log([\mu-V_{\rm min}]/\hslash\omega_\perp)\approx\mbox{const} \gg 1$
is the large logarithm. The corresponding variational equation of motion 
\begin{equation}
\Gamma[{\bf R}_\beta\times{\bf R}_t]\rho_0({\bf R})=\delta {\cal H}/\delta{\bf R},
\end{equation}
after its resolution with respect to the temporal derivative, looks as follows \cite{SF2000,R2001,R2016}:
\begin{equation}\label{LIA}
{\bf R}_t =\frac{\Gamma\Lambda}{4\pi} \Big\{\kappa {\bf b} 
+\Big[\frac{\nabla \rho_0({\bf R})}{\rho_0({\bf R})}\times \frac{{\bf R}_\beta}{|{\bf R}_\beta|}\Big]
\Big\} +{\bf v}_0({\bf R}),
\end{equation}
where $\kappa$ is the local curvature of the filament, ${\bf b}$ is the unit binormal vector, and  
${\bf R}_\beta/|{\bf R}_\beta|$ is the unit tangent vector.
Solutions of this equation were investigated for harmonic 3D traps 
(mainly --- in linearized form; see, e. g., \cite{Kelvin_vaves,ring_istability}) 
or for strictly 2D density profiles \cite{R2016}. To the best author's knowledge, no essentially
non-stationary exact solutions were reported so far. In this work some exact, finite-dimensional
integrable reduction of Eq.(\ref{LIA}) will be presented for the case of anharmonic trap
with a Gaussian profile of the equilibrium condensate density. The Gaussianity is supposed 
in that region of space where Eq.(\ref{LIA}) is applicable, and where the main energetics 
of the vortex is concentrated, in essence. Deviation from Gaussianity  is implied closely to
the Thomas-Fermi surface, but in the main approximation it should not be taken into account.

\vspace{2mm}

{\bf Simplification in the case of Gaussian density}.
The choice of the equilibrium density profile in the form 
$\rho_0({\bf r})\propto\exp(-{\bf r}\cdot\hat D {\bf r})$, with some constant positively defined symmetric
matrix $\hat D$, is dictated, first, by the fact that the logarithm gradient coming to Eq.(\ref{LIA})
is in this case $-2\hat D{\bf r}$, i. e., it is linear on ${\bf r}$. Second, for density profiles
depending on combination ${\bf r}\cdot\hat D {\bf r}$ only, equations (\ref{stat}) have a simple explicit
solution, and the velocity field  ${\bf v}_0$ is linear on ${\bf r}$ as well:
\begin{equation}\label{v_0}
{\bf v}_0=-[{\bf B}\times\hat D{\bf r}],\qquad {\bf B}=
2(\hat I \mbox{Tr\,} \hat D -\hat D)^{-1}{\bf \Omega},
\end{equation}
where $\hat I$ is the unit matrix $3\times 3$. Indeed, having supposed solution in the form 
${\bf v}_0=\hat A {\bf r}$, after substitution it into equations (\ref{stat}) we have the system of
algebraic equations for matrix $\hat A$:
\begin{equation}
\epsilon_{ijk}A_{jk}=2\Omega_i, \quad \mbox{Tr\,}\hat A=0,\quad \hat D\hat A +\hat A^T\hat D=0.
\end{equation}
Writing these equations for each component in the basis where $\hat D$ is diagonal, we find
\begin{equation}
A_{ij}=2\epsilon_{ijk}\Omega_k\frac{d_j}{d_i+d_j},
\end{equation}
where $d_i>0$ are eigenvalues of matrix $\hat D$. Let us note now that
\begin{equation}
2\epsilon_{ijk}\frac{\Omega_k}{d_i+d_j}=\epsilon_{ijk}B_k, \qquad 
B_k=\frac{2\Omega_k}{ -d_k + \sum_j d_j },
\end{equation}
so expressions (\ref{v_0}) follow from here.

Taking into account the formulas above, the local induction equation (appropriately non-dimensionalized)
in the Gaussian case looks as follows:
\begin{equation}\label{LIA_Gauss}
{\bf R}_t =\kappa {\bf b} +\Big[\frac{{\bf R}_\beta}{|{\bf R}_\beta|}
\times \hat D{\bf R}\Big] -[{\bf B}\times\hat D{\bf R}].
\end{equation}

\vspace{2mm}

{\bf Integrable reduction}. The interesting observation is that Eq.(\ref{LIA_Gauss}) 
admits solutions in the form of a straight non-stationary vortex,
\begin{equation}\label{str_line}
{\bf R}(\beta,t)=\beta {\bf M}(t) +{\bf N}(t).
\end{equation}
The vortex line curvature is identically zero in this case, and after substitution 
(\ref{str_line}) into (\ref{LIA_Gauss}) we obtain the system of ordinary differential
equations:
\begin{eqnarray}
\dot{\bf N}&=&\Big[\Big(\frac{\bf M}{|{\bf M}|}-{\bf B}\Big)\times\hat D{\bf N}\Big],
\label{N_eq}\\
\dot{\bf M}&=&\Big[\Big(\frac{\bf M}{|{\bf M}|}-{\bf B}\Big)\times\hat D{\bf M}\Big].
\label{M_eq}
\end{eqnarray}
It is easy to see the following integrals of motion:
\begin{equation}
{\bf N}\cdot \hat D {\bf N}=C_1, \quad {\bf M}\cdot \hat D {\bf N}=C_0,
\quad {\bf M}\cdot \hat D {\bf M}=C_2.
\end{equation}
Without loss of generality one can put  $C_0=0$. Then it becomes clear that at every time moment
the straight vortex touches the ellipsoid ${\bf r}\cdot \hat D {\bf r}=C_1$, with ${\bf N}$ being
the touching point.

Now we note that Eq.(\ref{M_eq}) for the tangent vector  ${\bf M}$ possesses a non-canonical 
Hamiltonian structure, which is similar to the structure of Landau-Lifshits equation, 
but with participation of matrix $\hat D$:
\begin{equation}
\dot{\bf M}=\left[\frac{\partial H({\bf M})}{\partial{\bf M}}\times\hat D{\bf M}\right],\qquad
H=|{\bf M}|-{\bf B}\cdot{\bf M}.
\end{equation}
Up to this moment, it was not assumed that the angular velocity vector is time-independent
in the trap-co-rotating system: the above equations are correct for non-stationary ${\bf\Omega}(t)$
as well. But if ${\bf\Omega}=$ const (as it will be accepted below), then the ``Hamiltonian''  
$H({\bf M})=C$ provides one more integral of motion.

The length of vector ${\bf M}$ does not have an immediate geometrical meaning, but the direction 
${\bf m}={\bf M}/|{\bf M}|$ is only important. Therefore, for further investigation of properties of
the dynamical system, it is convenient to use the combination of conservation laws which contains
the unit tangent vector ${\bf m}$ only:
\begin{equation}
\gamma({\bf m})\equiv\frac{1-{\bf B}\cdot{\bf m}}{\sqrt{{\bf m}\cdot \hat D {\bf m}}}=\mbox{const}, 
\qquad {\bf m}^2=1.
\end{equation}
It is interesting to note that equation of motion for ${\bf m}$ has a somewhat different structure:
\begin{equation}
\dot{\bf m}=\sqrt{({\bf m}\cdot \hat D {\bf m})^3}
\Big[\frac{\partial \gamma({\bf m})}{\partial{\bf m}}\times{\bf m}\Big].
\end{equation}
Trajectories of the system are level contours of function $\gamma({\bf m})$ at the unit sphere.
Depending on relations between the quantities $d_i$ and on the vector parameter ${\bf B}$,
phase portraits can be qualitatively different. Let us introduce the parametrization
\begin{equation}
d_1=1+\alpha,\qquad d_2=1-\alpha,\qquad d_3=\lambda,
\end{equation}
\begin{equation}
{\bf m}=(\sqrt{1-q^2}\cos\phi, \sqrt{1-q^2}\sin\phi, q),
\end{equation}
(where $q$ is cosine of the polar angle), and consider some examples of how the ``controlling parameter''
${\bf \Omega}$ changes the vortex dynamics.

\begin{figure}
\begin{center}
 (a) \epsfig{file=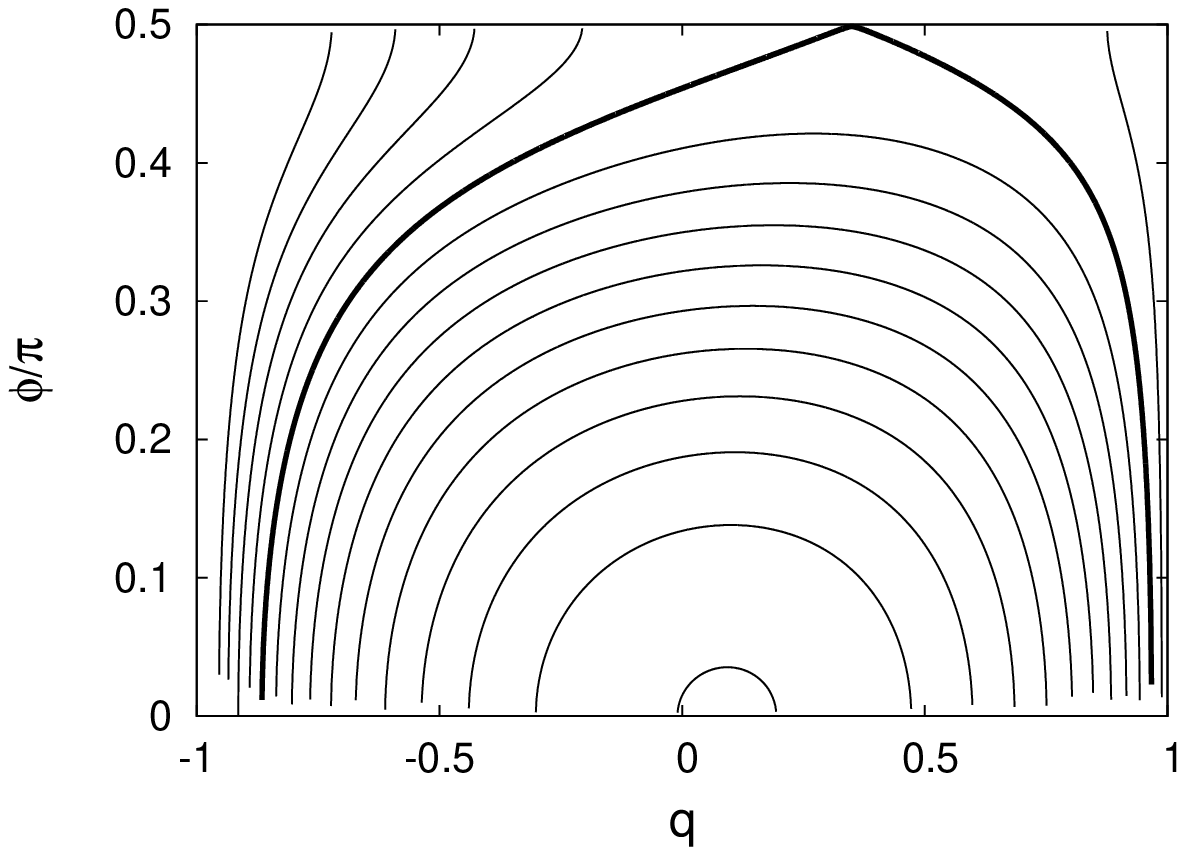, width=70mm}\\
\vspace{2mm}
 (b) \epsfig{file=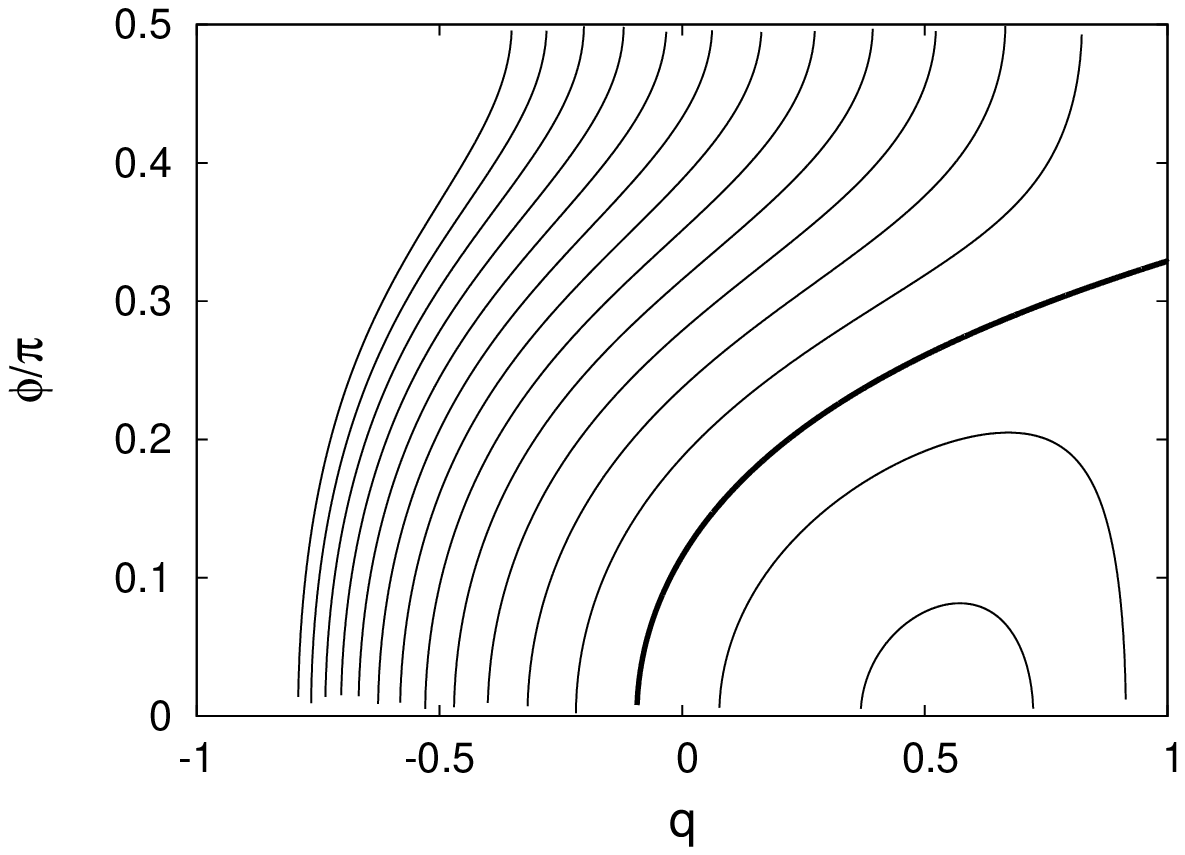, width=70mm}
\end{center}
\caption{ Formula (\ref{exA}) for $\alpha=0.3$, $\lambda =0.6$; a) $\Omega=0.05$, b) $\Omega=0.3$.
Bold lines indicate separatrixes.}
\label{ex1} 
\end{figure}
\begin{figure}
\begin{center}
  (a) \epsfig{file=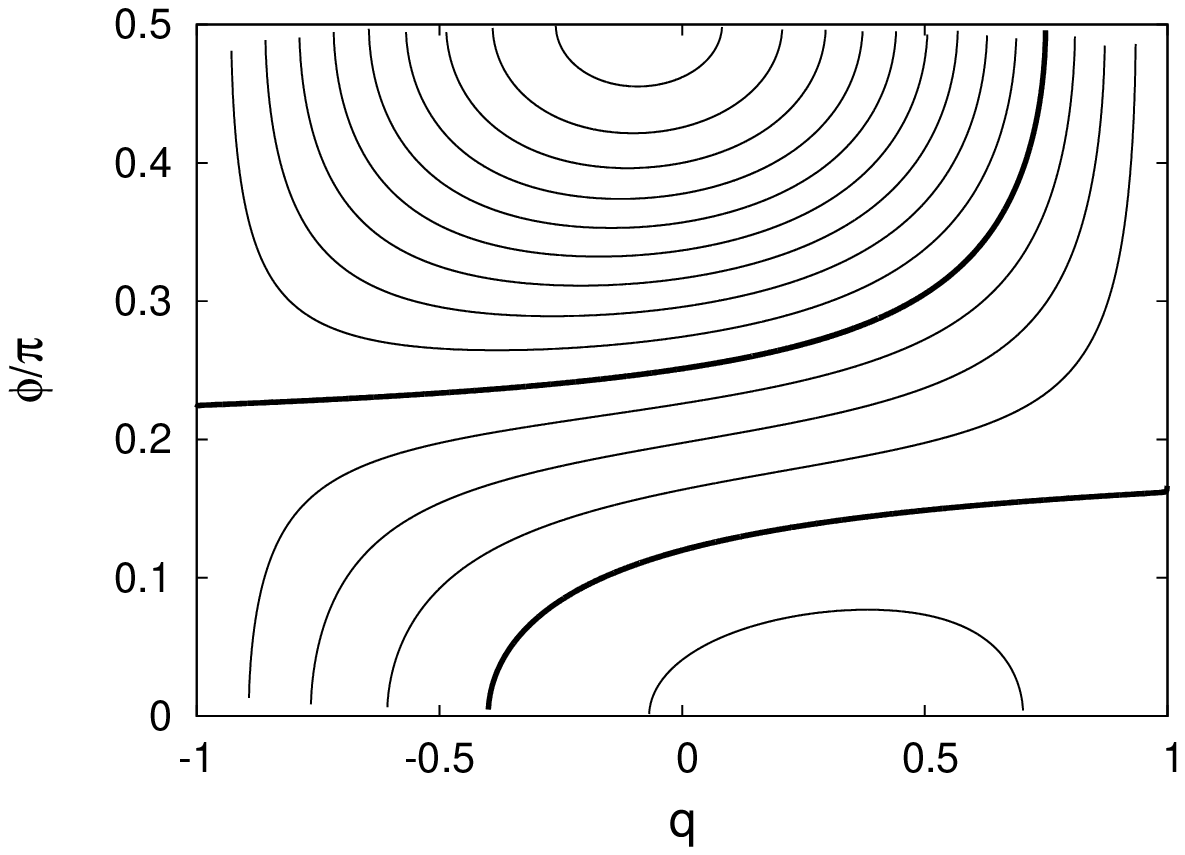, width=70mm}\\
\vspace{2mm}
  (b) \epsfig{file=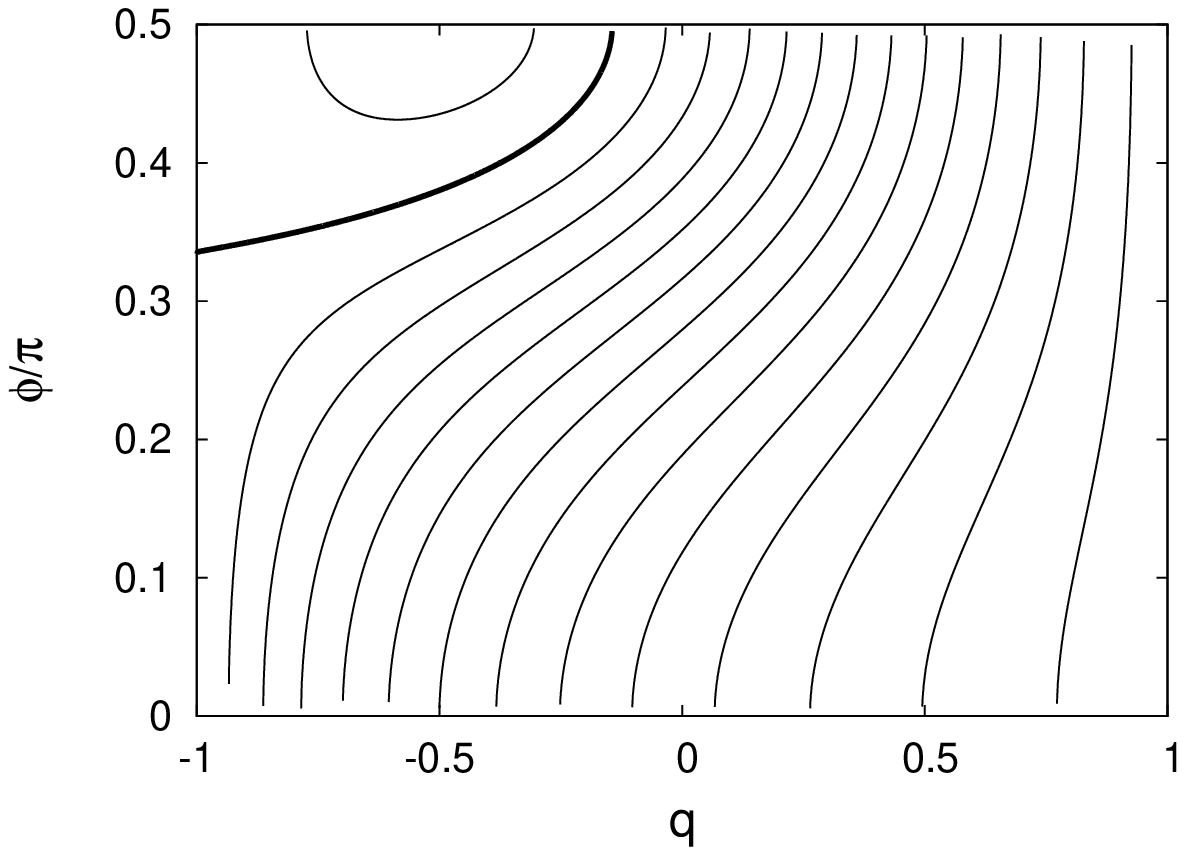, width=70mm}
\end{center}
\caption{ Formula (\ref{exA}) for $\alpha=0.3$, $\lambda =1.1$;  a) $\Omega=0.05$, b) $\Omega=0.3$.}
\label{ex2} 
\end{figure}
\begin{figure}
\begin{center}
 (a) \epsfig{file=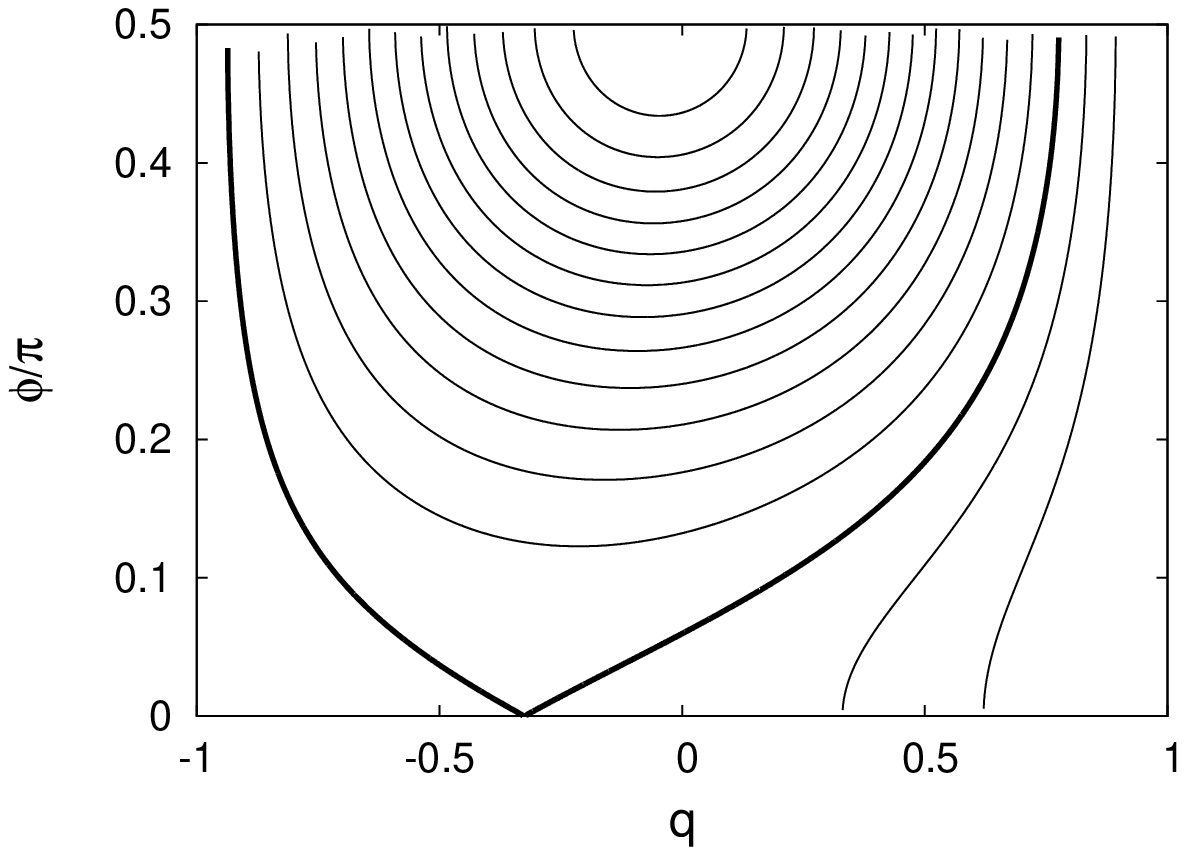, width=70mm}\\
\vspace{2mm}
 (b) \epsfig{file=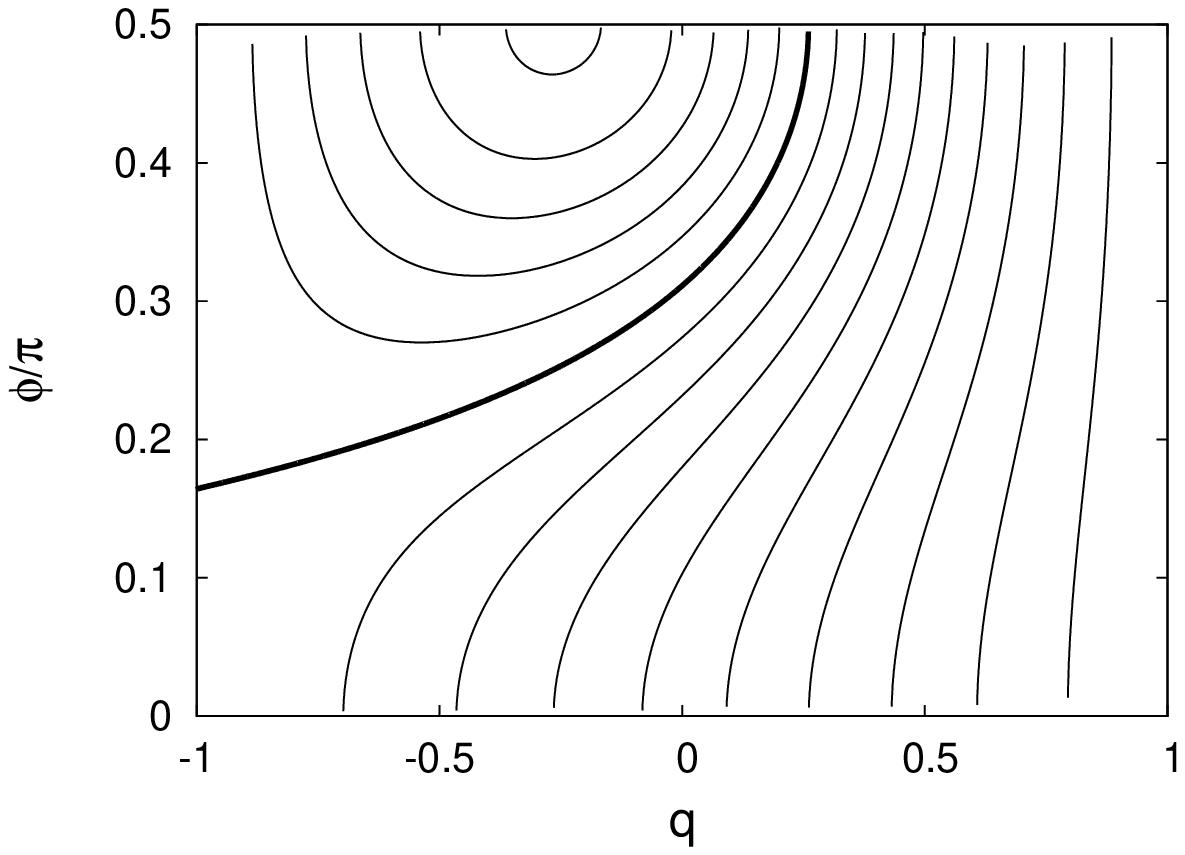, width=70mm}
\end{center}
\caption{ Formula (\ref{exA}) for $\alpha=0.3$, $\lambda =1.5$;  a) $\Omega=0.05$, b) $\Omega=0.3$.}
\label{ex3} 
\end{figure}

\vspace{2mm}

{\bf Example 1}. Let vector  ${\bf\Omega}$ be directed along $z$ axis. Then we have the family 
of curves mutually distinguished by value of parameter $\gamma$:
\begin{equation}\label{exA}
1-\Omega q=\gamma\sqrt{(1-q^2)(1+\alpha\cos 2\phi)+\lambda q^2}.
\end{equation}

In Fig.1, shown are phase portraits at $\alpha=0.3$, $\lambda =0.6$  (rotation around the largest
axis of a three-axial ellipsoid) for $\Omega=0.05$ and for $\Omega=0.3$ (a quarter of the unit sphere
is only shown; the curves should be mentally continued by symmetry in the azimuthal direction).
It is seen that at slower rotation, both ``poles'' are stable singular points of ``center'' type.
At  $x$ meridians there two more centers, and at $y$ meridians there are two ``saddle'' points.
Such a structure of integral lines seems natural, because at zero rotation frequency we deal essentially 
with the classical equations describing the motion of a rigid body. At faster rotation the ``South Pole''
remains to be a center, while the two meridional saddle points approach the ``North Pole'' and 
transform it by bifurcation into a saddle.

In Fig.2, the parameter  $\lambda =1.1$, i.e. the rotation takes place around the middle axis of 
an ellipsoid. At small angular velocity both poles are saddle points,  two centers are located
at $x$ meridians, and two more centers are located at $y$ meridians. With increase of $\Omega$,
the centers at $x$ meridians approach the North Pole and transform it to a center.

In Fig.3, the parameter   $\lambda =1.5$, i.e. the rotation occurs around the small axis. In this case,
as  $\Omega$ increases, two saddle points at $x$ meridians approach the originally stable South Pole 
and transform it to a saddle, while the North Pole remains to be a center, and two more centers continue 
to exist at $y$ meridians.

It should be said that with even faster rotation, in all the cases there remain two centers only (not shown).
But it is necessary to keep in mind that in reality at fast rotation two or more mutually interacting
vortices penetrate into the condensate.

\begin{figure}
\begin{center}
 (a)  \epsfig{file=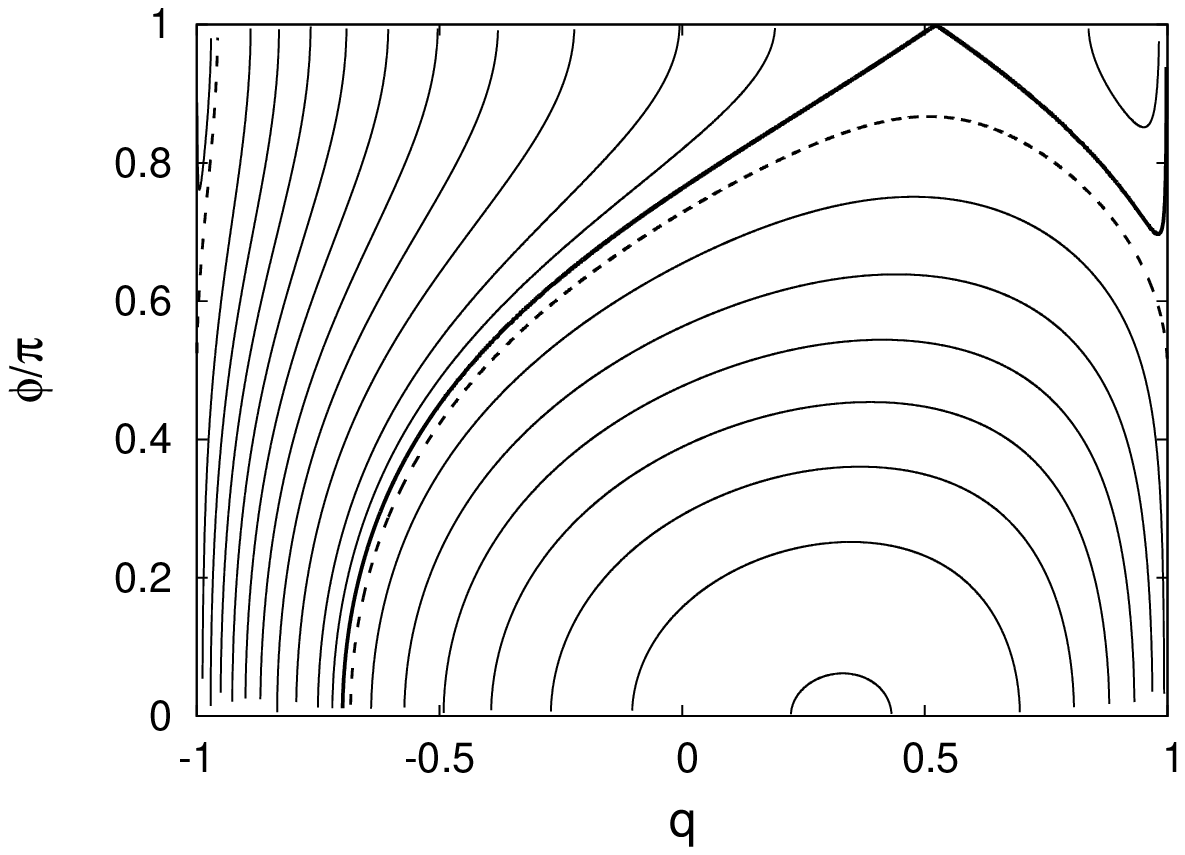, width=70mm}\\
\vspace{2mm}
 (b) \epsfig{file=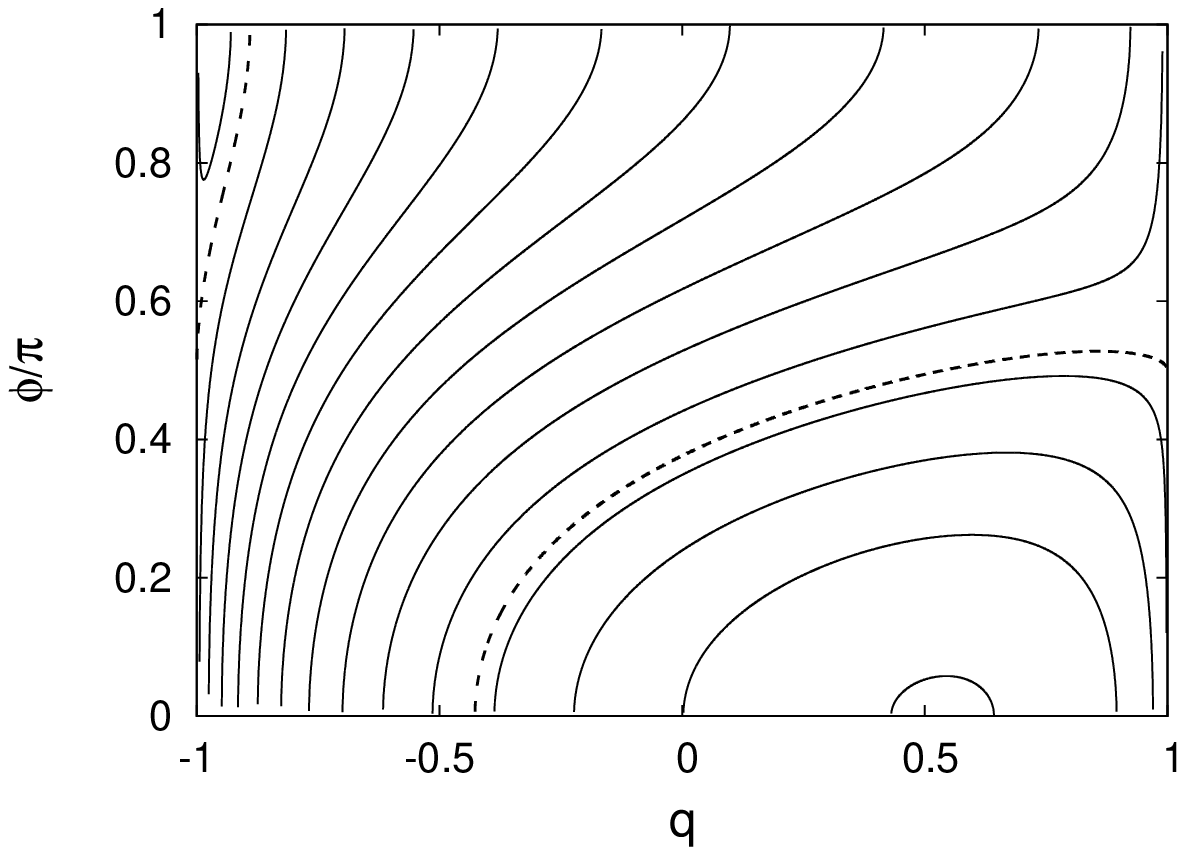, width=70mm}
\end{center}
\caption{ Formula (\ref{exB}) for $\lambda =0.5$; a) $B_3=B_1=0.2$; b) $B_3=B_1=0.4$.
Dashed lines indicate non-singular trajectories passing through the points $\pm{\bf e}_z$.}
\label{ex4} 
\end{figure}
\begin{figure}
\begin{center}
 (a)  \epsfig{file=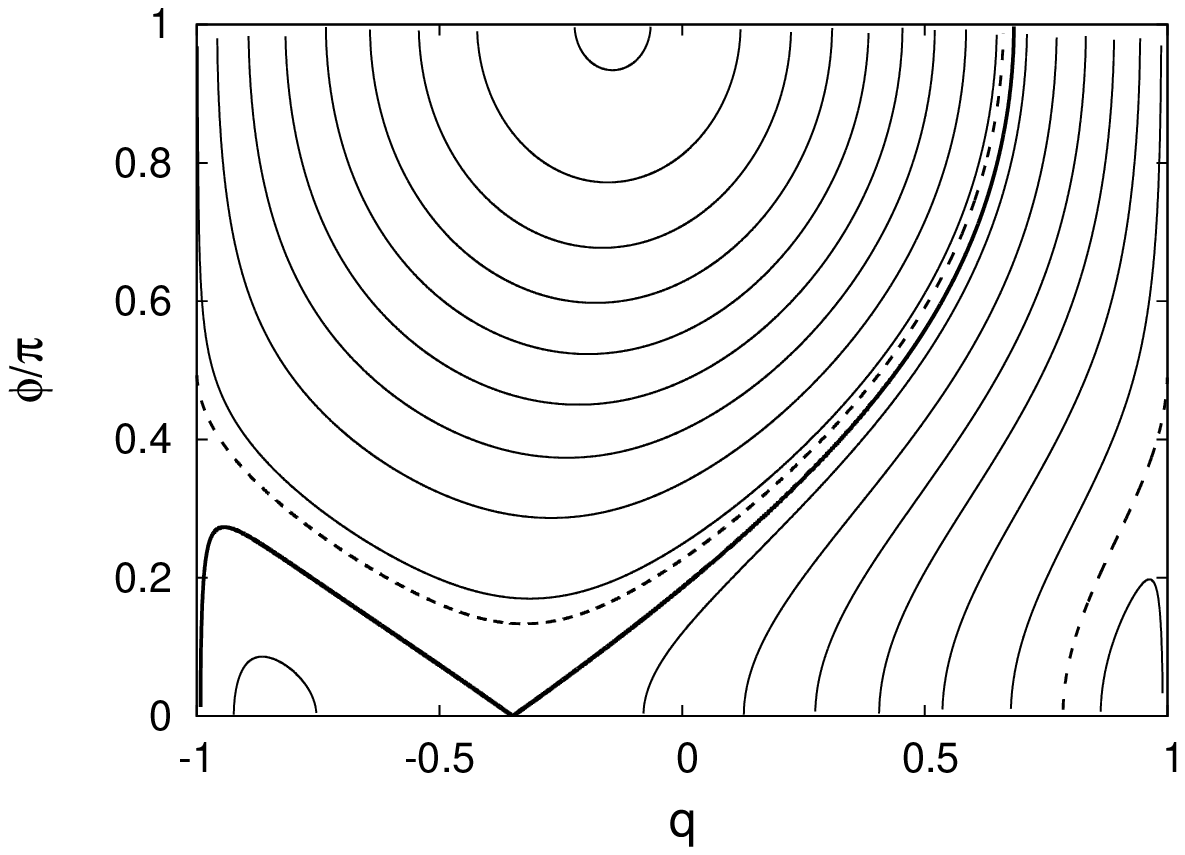, width=70mm}\\
\vspace{2mm}
 (b)  \epsfig{file=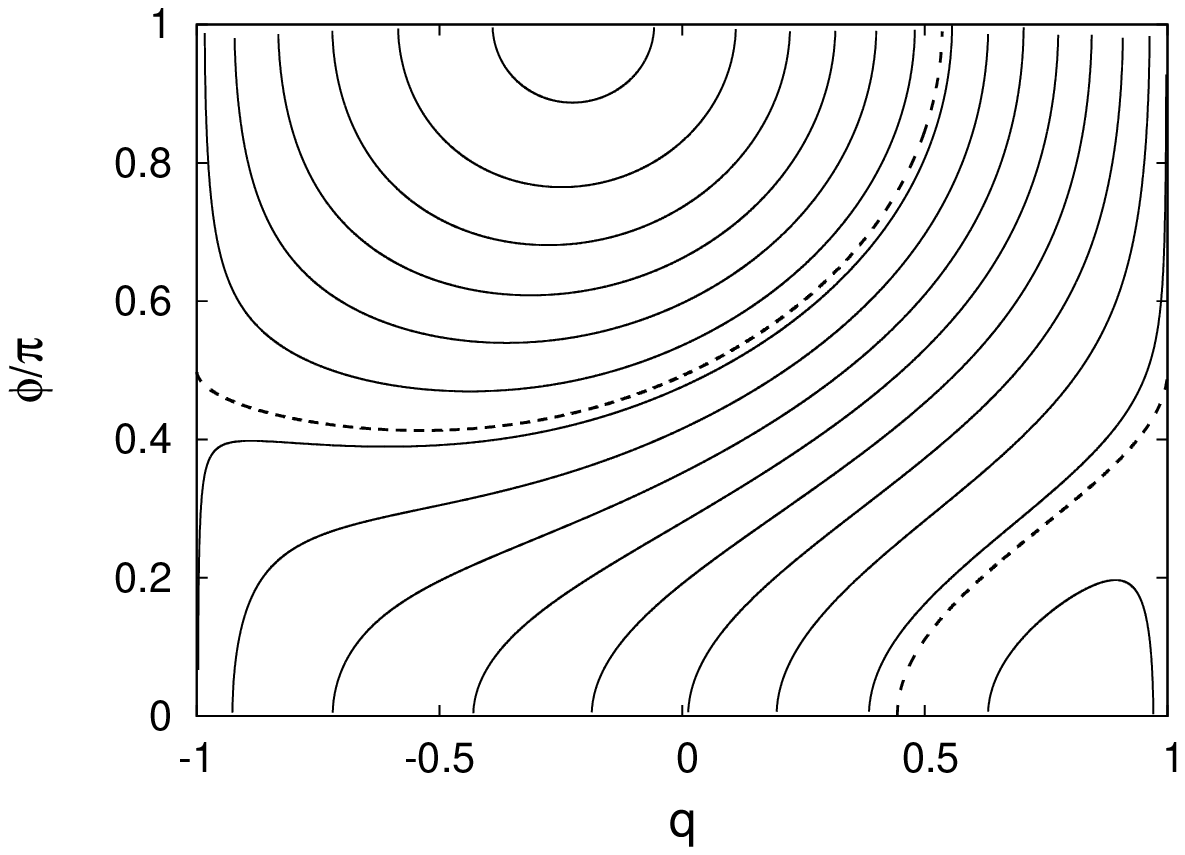, width=70mm}
\end{center}
\caption{ Formula (\ref{exB}) for $\lambda =2.0$; a) $B_3=B_1=0.2$; b) $B_3=B_1=0.4$.}
\label{ex5} 
\end{figure}

\vspace{2mm}

{\bf Example 2}. Let now the anisotropy  $\alpha =0$, but the vector ${\bf B}$ is oriented at some angle
to the symmetry axis of the ellipsoid. Then
\begin{equation}\label{exB}
1-B_3 q -B_1\sqrt{1-q^2}\cos\phi=\gamma\sqrt{1+(\lambda-1) q^2}.
\end{equation}

In Fig.4, shown are phase portraits in the case of a cigar-shaped axisymmetric ellipsoid with
$\lambda =0.5$ for two co-oriented vectors ${\bf B}$, differing by absolute values (only a half of 
the unit sphere is shown; the curves should be continued by symmetry in the azimuthal direction).
At smaller rotation frequency there are two centers at $x_-$ meridian closely to  $\pm{\bf e}_z$,
and one more center at $x_+$ meridian, while at  $x_-$ meridian there is also a saddle. At faster rotation
the saddle and center near the North Pole mutually delete each other, so only two centers finally remain
on the sphere.

In Fig.5, shown is the case of a disk-shaped axisymmetric ellipsoid with $\lambda =2.0$. Here at slow 
rotation there is one center at $x_-$ meridian, two centers near the poles at $x_+$ meridian, and
also a saddle at $x_+$ meridian. At faster rotation, annihilation of saddle and center near the 
South Pole occurs.

\vspace{2mm}

{\bf Conclusions}.  Thus, in this work it has been theoretically shown that with Gaussian density 
background, the 3D dynamics of a single vortex filament in rotating, essentially anisotropic Bose-Einstein
condensate can occur in the regime of straight off-center vortex, when the trap rotation in combination 
with spatial non-uniformity dominate over the line curvature effect. The corresponding integrable system
of ordinary differential equations has been analyzed. By changing the rotation speed of the trap, one
can to some extent manipulate the behaviour of straight vortex. It is naturally to suggest that even 
if non-local corrections to equation of vortex motion are taken into account and/or in close-to-Gaussian
cases, a qualitatively similar regime is possible, when curvature of the vortex line is non-small near 
the condensate surface only. At least, for an almost-spherical non-rotating harmonic trap, approximate
solutions in the form of a straight vortex passing the origin, were found in \cite{SF2000}.

\end{document}